\documentclass[aps,superscriptaddress,showpacs,twocolumn,floatfix,amsfonts]{revtex4-1}
\usepackage[dvips]{graphicx}

\usepackage{color}

\begin{document}

\title{Nonstationary regime for quasinormal modes of the charged Vaidya metric}

\def \beq {\begin{equation}}
\def \eeq {\end{equation}}

\author{Cecilia Chirenti}\email{cecilia.chirenti@ufabc.edu.br}
\affiliation{Centro de Matem\'atica, Computa\c c\~ao e Cogni\c c\~ao, UFABC, 09210-170 Santo Andr\'e, SP, Brazil}
\author{Alberto Saa}\email{asaa@ime.unicamp.br}
\affiliation{Departamento de Matem\'atica Aplicada, UNICAMP,  13083-859 Campinas, SP, Brazil}

\date{\today}

\begin{abstract}
We consider non-stationary spherically symmetric $n$-dimensional charged black holes
with varying mass $m(v)$ and/or electric charge $q(v)$, described by generic charged Vaidya metrics with cosmological constant $\Lambda$
in double null coordinates, and  perform a comprehensive numerical analysis
of the fundamental quasinormal modes (QNM) for  minimally coupled scalar fields.
We show that the ``instantaneous'' quasinormal frequencies exhibit the same sort of non-stationary behavior reported previously for the four-dimensional uncharged case with $\Lambda=0$.
 Such property seems to be very robust, independent of the spacetime dimension and of the metric parameters,
 provided they be consistent with the existence of an event horizon.
 The study of time dependent Reissner-Nordstr\"om black holes
allows us to go a step further and quantify the deviation of the
stationary regime for QNM with respect to charge variations as well.
 We also look for signatures in the quasinormal frequencies from the
creation of a Reissner-Nordstr\"om  naked spacetime singularity.
Even though one should expect the breakdown of our
approach in the presence of   naked singularities, we show that it is   possible, in principle, to
obtain some information about the naked singularity from the QNM frequencies, in agreement with the previous
results of Ishibashi and Hosoya  showing that it would be
indeed possible   to have regular scattering from naked singularities.
\end{abstract}


\maketitle

\section{Introduction}

The Quasinormal Modes (QNM) analysis is a central tool  in the
investigation  of gravitational  perturbations of stars and black
holes. (For recent comprehensive reviews of the vast literature on the
subject, see \cite{Berti:2009kk} and \cite{Konoplya:2011qq}.) The QNM
analysis of black holes is particularly relevant due to the
possibility of observation of the ringdown signals in gravitational
wave detectors such as LISA, see  \cite{Berti:2005ys} for
instance. Besides their relevance for future observations, the  QNM can also reveal important information about the
structure and behavior of  solutions of the Einstein equations. Many
generalized solutions have been investigated, including
non-asymptotically flat \cite{Brady:1996za,Molina:2003dc,Du:2004jt},
charged (see \cite{Konoplya:2007jv} for a comprehensive analysis), and
time-dependent ones
\cite{Hod:2002gb,Xue:2003vs,Shao:2004ws,Abdalla:2006vb,Abdalla:2007hg,CuadrosMelgar:2007bm,He:2009jd}.
The AdS/CFT conjecture, in particular, has motivated many QNM analyses
of generalized black hole solutions \cite{Wang:2005vs}. According to the conjecture, the QNM frequencies carry
information about the thermal properties of the associated conformal
field theory \cite{Berti:2009kk}.

The Vaidya metric \cite{Kramer} was originally proposed to describe the spacetime outside a radiating star.
It has also been the usual starting point for
the study of time dependent black holes QNM
\cite{Xue:2003vs,Shao:2004ws,Abdalla:2006vb,Abdalla:2007hg,He:2009jd}.
Namely, it corresponds to a time dependent solution of  Einstein
equations for a  spherically symmetric body immersed in   a
unidirectional radial null-fluid flow. It has been also widely used in
the analysis of spherically symmetric collapse and the formation of
naked singularities for many years. (For further references, see
\cite{Joshi} and the extensive list of \cite{Lake}). It is also  known
that the Vaidya metric can be obtained from the Tolman metric by
taking appropriate limits in the self-similar
case\cite{LemosHellaby}. This result has shed some light on the nature
of the so-called shell-focusing singularities \cite{EardleySmarr}, as
discussed in detail in
\cite{Lake,LemosHellaby,EardleySmarr,Kuroda1,LakeZannias1,LakeZannias2}.
The Vaidya metric has also proved to be useful in the study of Hawking
radiation and the process of black-hole evaporation
\cite{Hiscock,Kuroda,Biernacki,Parentani}, in the stochastic gravity
program \cite{Bei-Lok}, and in recent numerical relativity
investigations \cite{Nielsen:2010wq}. The charged version of the
metric is also an usual starting point to the study of many  aspects
of charged black hole physics and naked singularities
\cite{charge1,charge2,Ori91,Fayos,Parikh:1998ux}.

The main purpose of the present paper is to consider the QNM of time
dependent backgrounds corresponding to a general Vaidya metric, with
special emphasis on nonstationary effects, generalizing and further developing the work done in \cite{Abdalla:2006vb,Abdalla:2007hg}. This kind of  QNM analysis
is certainly   relevant from the physical point of view since, for
instance, it is quite natural to expect that a real black hole be
affected by processes which can change its mass such as, for instance,
mass accretion  or even Hawking radiation, which would indeed imply a
decreasing mass. Any signal coming from a black hole could, in principle,
have some nonstationary component.

As a secondary objective, we show that our numerical setup can be used
to investigate the QNM frequencies in the case of a Vaidya metric
evolving towards a Reisser-Nordstr\"om naked singularity. One should
expect the breakdown of our approach in the presence of a naked
singularity, of course. However, we show that it is in principle possible to
obtain some information about the singularity from the QNM
frequencies, in agreement with the results of Ishibashi and Hosoya
\cite{Ishibashi:1999vw}, who conclude that it is indeed possible to have
regular scattering from naked singularities.

This paper has three more sections. In the next one we present the main equations describing the most generic Vaidya metric in double-null
 coordinates and write the scalar wave equation in a canonical hyperbolic form, specially
 suitable for the numerical analysis presented in Section III. In the
 last section, we present some concluding remarks.

\section{The scalar wave equation in the Vaidya spacetime}
\label{sec2}

The $n$-dimensional Vaidya metric was first discussed in \cite{IV}. It
can be easily cast  in $n$-dimensional radiation coordinates
$(v,r,\theta_1,\dots,\theta_{n-2})$ as done, for instance, in
\cite{GD}. The $n$-dimensional charged Vaidya metric in radiation
coordinates, obtained originally in \cite{charge2}, reads
\begin{eqnarray}
\label{Vaidya}
ds^2 &=& -\left(1-\frac{2m(v)}{(n-3)r^{n-3}}
+ \frac{q^2}{(n-2)(n-3)r^{2(n-3)}}
\right)dv^2\nonumber\\ && +2cdrdv+ r^2d\Omega^2_{n-2},
\end{eqnarray}
 where $n>3$,  $c=\pm 1$, and $d\Omega^2_{n-2}$ stands for the metric
 of the unit $(n-2)$-dimensional sphere, assumed here to be spanned by
 the angular coordinates $(\theta_1,  \theta_{2},\dots
 ,\theta_{n-2})$ in the usual way.
For the case of an ingoing radial flow, $c=1$ and $m(v)$ is a monotonically
increasing mass function in the advanced time $v$, while $c=-1$
corresponds to an outgoing radial flow, with $m(v)$ being in this case
a monotonically decreasing mass function in the retarded time $v$. The
constant $q$ corresponds to the total electric charge. In principle,
one can also consider time dependent charges $q$ as done, for instance,
in \cite{Ori91}. This situation will of course require the presence of
charged null fluids and currents, whose realistic nature  we do not
address here. We have already reported some preliminary results on QNM
for this case in \cite{Chirenti:2010iu}.

We will now extend the approach proposed in \cite{GS,Saa:2007ej}
 and derive the double-null formulation for the most general Vaidya
 metric:  $n$-dimensional, in the presence of a cosmological constant,
 and with  varying electric charge.  Only the main results are
 presented. The reader can get more details on the employed
 semi-analytical approach in \cite{GS,Saa:2007ej} and the references
 cited therein. We recall that the $n$-dimensional spherically
 symmetric line element in double-null coordinates
 $(u,v,\theta_1,\dots,\theta_{n-2})$ is given by
\beq
\label{uv}
ds^2 = -2f(u,v)du\,dv + r^2(u,v)d\Omega^2_{n-2},
\eeq
where $f(u,v)$ and $r(u,v)$ are non vanishing smooth functions. The
energy-momentum tensor of a unidirectional radial null-fluid in the
eikonal approximation in the presence of an electromagnetic field
$F_{ab}$ is given by
\beq
\label{T}
T_{ab} = \frac{1}{8\pi}h(u,v)k_a k_b
+\frac{1}{4\pi}\left(F_{ac}F^{\phantom{b}c}_{b} -\frac{1}{4}g_{ab}
F_{cd}F^{cd}\right)
,
\eeq
where $k_a$ is a radial null vector and $h(u,v)$ is a smooth function
characterizing the null-fluid radial flow. We will consider here,
without loss of generality, the case of a flow along the $v$-direction.

From the Einstein-Maxwell equations with metric (\ref{uv}) and energy-momentum tensor (\ref{T}) we obtain the following equations for the functions $f$, $h$ and $r$:
\beq
\label{B}
f = 2Br_{,u},
\eeq
\beq
\label{h}
h = -2\left(\frac{n-2}{n-3}\right)  \frac{B}{r^{n-2}}
\left(m_{,v} - \frac{1}{n-2}\frac{(q^2)_{,v}}{r^{n-3}} \right),
\eeq
%
\begin{eqnarray}
\label{r2}
r_{,v} &=& -B \left(1 - \frac{2m(v)}{(n-3)r^{n-3}} - \frac{2\Lambda r^2}{(n-2)(n-1)} + \right.\nonumber\\
&+& \left. \frac{2q^2(v)}{(n-2)(n-3)r^{2(n-3)}}
\right),
\end{eqnarray}
where $B(v)$, $m(v)$ and $q(v)$ are arbitrary integration functions. If we choose $B(v) = $ constant, we can interpret $m(v)$ and $q(v)$ as the mass and charge of the solution, respectively. These two functions must be monotonic, and must be chosen in a way that satisfies the null-energy condition \cite{Ori91}. The details of the derivation of eqs.(\ref{B})-(\ref{r2}) and the full analysis of the charged Vaidya metric in duble null coordinates will be presented in another paper currently under preparation.

For our QNM analysis we will consider the evolution of a  massless scalar field governed by
the Klein-Gordon equation
\beq
\frac{1}{\sqrt{-g}}\left(\sqrt{-g}g^{ab}\Psi_{,b}\right)_{,a}
= 0,
\label{wave1}
\eeq
in the background (\ref{uv}). Taking advantage of the spherical
symmetry, we decompose the scalar field in terms of higher-dimensional
spherical harmonics
\beq
\Psi  =
\sum_{\ell,m}\psi(u,v)Y_{\ell m}(\theta_1,\ldots,\theta_{n-2})\,,
\label{psi}
\eeq
for which
\beq
\nabla^2_{\Omega}Y_{\ell m} = -\ell(\ell + n -3)Y_{\ell m},
\eeq
where $\nabla^2_{\Omega}$ stands for the Laplacian operator over the
$(n-2)$-dimensional unit sphere. Substituting the ansatz (\ref{psi})
in (\ref{wave1}), we obtain for the metric (\ref{uv})
\beq
\frac{1}{fr^{n-2}}\left( (r^{n-2}\psi_{,v})_{,u} +
  (r^{n-2}\psi_{,u})_{,v}\right) + \frac{\ell(\ell+n-3)}{r^2}\psi\,.
\eeq
Using now the substitution $\psi = r^{-\frac{n-2}{2}}\varphi,$
we get
\beq
\varphi_{,uv} + V(u,v)\varphi = 0\,,
\label{phi}
\eeq
where
\beq
V=\frac{\ell(\ell+n-3)}{2r^2}f -
  \frac{(n-2)(n-4)}{4r^2}r_{,u}r_{,v}
   - \frac{(n-2)}{2r}r_{,uv} .
\label{V1}
\eeq

The wave equation (\ref{phi}) is already written in a canonical hyperbolic
form and it will be the starting point for our QNM
analysis. Nevertheless, the potential (\ref{V1}) can still be cast in a
more convenient way. From equations (\ref{B}) and (\ref{r2}), we find
\beq
r_{uv} = -f\left(\frac{m}{r^{n-2}} - \frac{2\Lambda r}{(n-2)(n-1)} -
  \frac{2q^2}{(n-2)r^{2n-5}} \right)\,,
\label{ruv}
\eeq
and
\begin{eqnarray}
r_{,u}r_{,v} &=& -f\left(\frac{1}{2} - \frac{m}{(n-3)r^{n-3}} -
  \frac{\Lambda r^2}{(n-2)(n-1)} \right. \nonumber \\
  && \left.
+ \frac{q^2}{(n-2)(n-3)r^{2(n-3)}}
  \right)\,.
\end{eqnarray}
Now we can finally write the potential $V(u,v)$ as
%
\begin{eqnarray}
& &V(u,v) = \frac{f}{2}\left(\frac{\ell(\ell+n-3)}{r^2} +
\frac{(n-2)(n-4)}{4r^2} + \right. \nonumber \\
&+& \left. \frac{(n-2)^2m(v)}{2(n-3)r^{n-1}}   -
\frac{n\Lambda}{2(n-1)} - \frac{(3n-8)q^2(v)}{2(n-3)r^{2(n-2)}}\right).
\label{potential}
\end{eqnarray}
%
The conventions we adopted for the mass, electric charge and
cosmological constant appearing in the potential (\ref{potential}) and
used in the derivations of this section are the standard ones
employed in the Vaidya metric literature. In particular, our
expressions coincide, in the appropriate limits, with the previous
results for the $\Lambda=0$ \cite{charge2} and for the $q=0$
\cite{Saa:2007ej} cases. However, the commonly employed conventions in
the QNM literature are slightly different, see, for instance,
\cite{Konoplya03}.

\section{Numerical results}

The wave equation (\ref{phi}) is already written in a canonical
hyperbolic form and it can be numerically integrated by means
of a characteristic scheme, {\em i.e.}, one can evolve $\varphi(u,v)$
in $v$ knowing $\varphi(u,v_0)$. In order to determine $V(u,v)$, one
needs to known $f(u,v)$ and $r(u,v)$ and, consequently, Eq. (\ref{r2})
must be solved in each evolution step. Since only $\varphi(u,v)$ is
required to determine $\varphi(u,v+dv)$, one can implement the
algorithm in an efficient way avoiding unnecessary calculations. We use
here the same numerical scheme  used in
\cite{Abdalla:2006vb,Abdalla:2007hg} to evolve the system and read the
QNM frequencies, which allows us to perform an exhaustive numerical
analysis with modest computational resources.

All mass functions considered in this work are of the form
\beq
\label{tanh}
2m(v) = \left(m_f+m_i\right) + \left(m_f-m_i\right)\tanh \rho_m (v-v_m),
\eeq
where $m_i$ and $m_f$ stand for the initial and final
 mass, respectively, and $\rho_m$ controls how fast the change is, with the maximum
 rate of change at $v=v_m$. We use an analogous expression
 for the time dependent electric charge $q(v)$. Our main results do
 not depend on the exact form of the functions $m(v)$ and
 $q(v)$. Choices such as (\ref{tanh}), however, are very convenient
 due to their smooth behavior and the ``asymptotically static" limits for $v\to \pm \infty$.

For the numerical calculation of the $v$-evolution, the algorithm
requires the evaluation of $\varphi(u,v)$  for all values of $u$
corresponding to the exterior region of the black hole, let us say,
for $u_h<u<\infty$, with $u=u_h$ corresponding to the event
horizon. Technically, it is easier to control the approach to the
regions near the horizon  by introducing an appropriated kind of tortoise coordinate
$U$ such that the external region   $u_h<u<\infty$ will correspond to
$-\infty<U<\infty$.

\subsection{Static case: Reissner-Nordstr\"om black hole}

It is  instructive to start by testing our code with the standard
  Reissner-Nordstr\"om (RN) case, for which a vast set of results is
  available in the literature \cite{Konoplya:2007jv}. The potential
  (\ref{potential}) for $n=4$ and $\Lambda=0$ becomes simply
\beq
V(u,v) = \frac{f}{2}\bigg(\frac{\ell(\ell+1)}{r^2}
+ \frac{2m}{r^{3}} -
 \frac{2q^2}{r^{4}}\bigg)\,.
\label{potential_RN}
\eeq
In order to compare the QNM frequencies obtained from the numerical
integration of (\ref{phi}) for the potential (\ref{potential_RN})
with well-known results for the RN black hole, we
perform the following change of variables
\beq
U = u - \frac{2r_+^2}{r_+ - r_-}\ln\left(-\frac{u}{2} - r_+ \right) -
\frac{2r_-^2}{r_+ - r_-}\ln\left(-\frac{u}{2} - r_- \right),
\label{U-RN}
\eeq
where
\beq
r_{\pm} = m \pm \sqrt{m^2 - q^2},
\eeq
as usual for the RN black hole. This is necessary in order to
guarantee that the QNM frequencies be defined with respect to the same
scales. Moreover, the new tortoise coordinate $U$ given by
(\ref{U-RN}) allows a better description of the exterior region of the
black hole, notably of the regions very close to the horizon
$r_+$. The numerical grid is, therefore, written in $U \times v$
coordinates, and $u$ is obtained by inverting the definition
(\ref{U-RN}).

\begin{figure}[!htb]
\begin{center}
  \includegraphics[angle=270,width=1.0\linewidth]{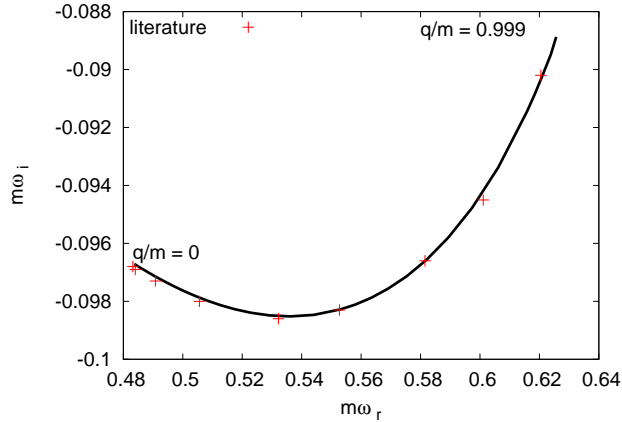}
\end{center}
\caption{Frequencies of the quasinormal modes of a scalar perturbation
  with $\ell = 2$ obtained for a RN black hole with our code (smooth
  line) compared with values found in the literature
  \cite{Konoplya03,Konoplya02}.}
\label{fig2b}
\end{figure}

A comparison
between the the numerical results obtained with our code and the
values found in the literature is shown in Fig.~\ref{fig2b}.
We observe a very good agreement with the known values for the RN QNM
 frequencies. This simple example allows us to calibrate all the
 algorithm parameters in order to attain a pre-established accuracy.

\subsection{Time-dependent case}

We perform a comprehensive QNM analysis for the wave equation
(\ref{phi}) with the potential (\ref{potential}). We are mainly
interested in the nonstationary regimes like the one described in
\cite{Abdalla:2006vb}, corresponding to the case $q=0$, $n=4$, and
$\Lambda=0$ in the potential (\ref{potential}). As we show in this section, the
stationary and nonstationary regimes for QNM are also present for the
general situation corresponding to the potential (\ref{potential}). Moreover,
we verify appreciable deviations from the stationary regime whenever
$|m''(v)|$ or $|q''(v)|$ is larger than $|\omega_I|$, where $\omega_I$
is the imaginary part of the first QNM frequency. This behavior occurs
irrespective of the other parameters of the potential
(\ref{potential}), provided they be consistent with the existence of a
black hole.

The case of an asymptotically flat ($\Lambda=0$) ``time dependent RN
black hole'', {\em i.e.}, a RN black hole with time dependent mass
$m(v)$ and/or charge $q(v)$ functions, is particularly
interesting. The determination of the new tortoise variable $U$
equivalent to (\ref{U-RN}) is quite more involved for this case. The
problem here is the definition, and the numerical determination, of
the event horizon. (See \cite{Saa:2007ej} for a
discussion of the implications of this problem in the semi-analytical
approach used here.)

For our purposes here,
the event horizon $r_+$ is the last null geodesic (up to the machine
precision) escaping towards infinity, requiring the full numerical
solution of (\ref{r2}) prior to the analysis of the wave equation
(\ref{phi}). The determination of the real Cauchy horizon $r_-$
is easier since any null geodesic inside the event horizon tends to
$r_-$ along the $v$-evolution.

\begin{figure*}[!htb]
  \includegraphics[angle=270,width=0.45\linewidth]{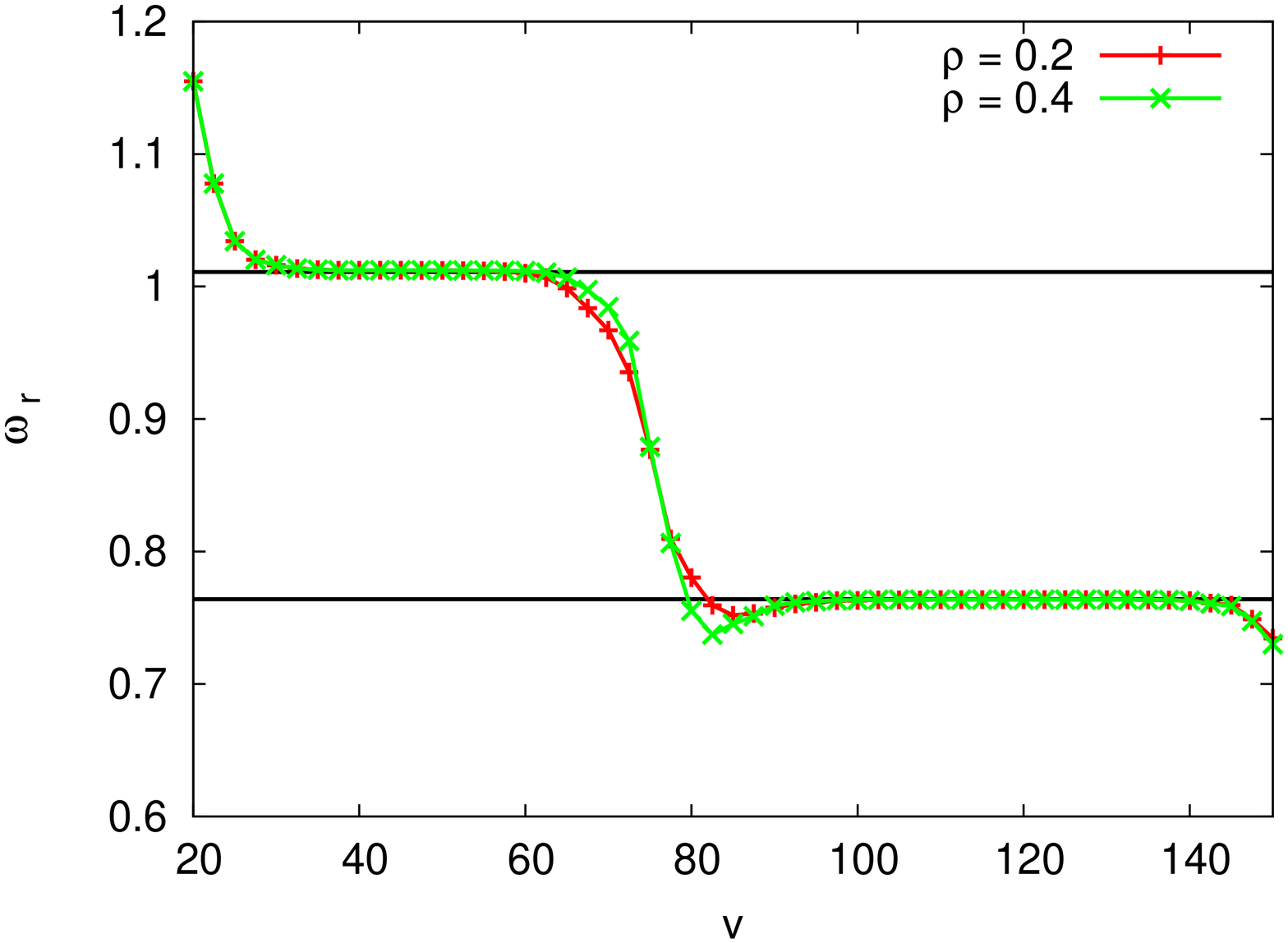}
  \includegraphics[angle=270,width=0.45\linewidth]{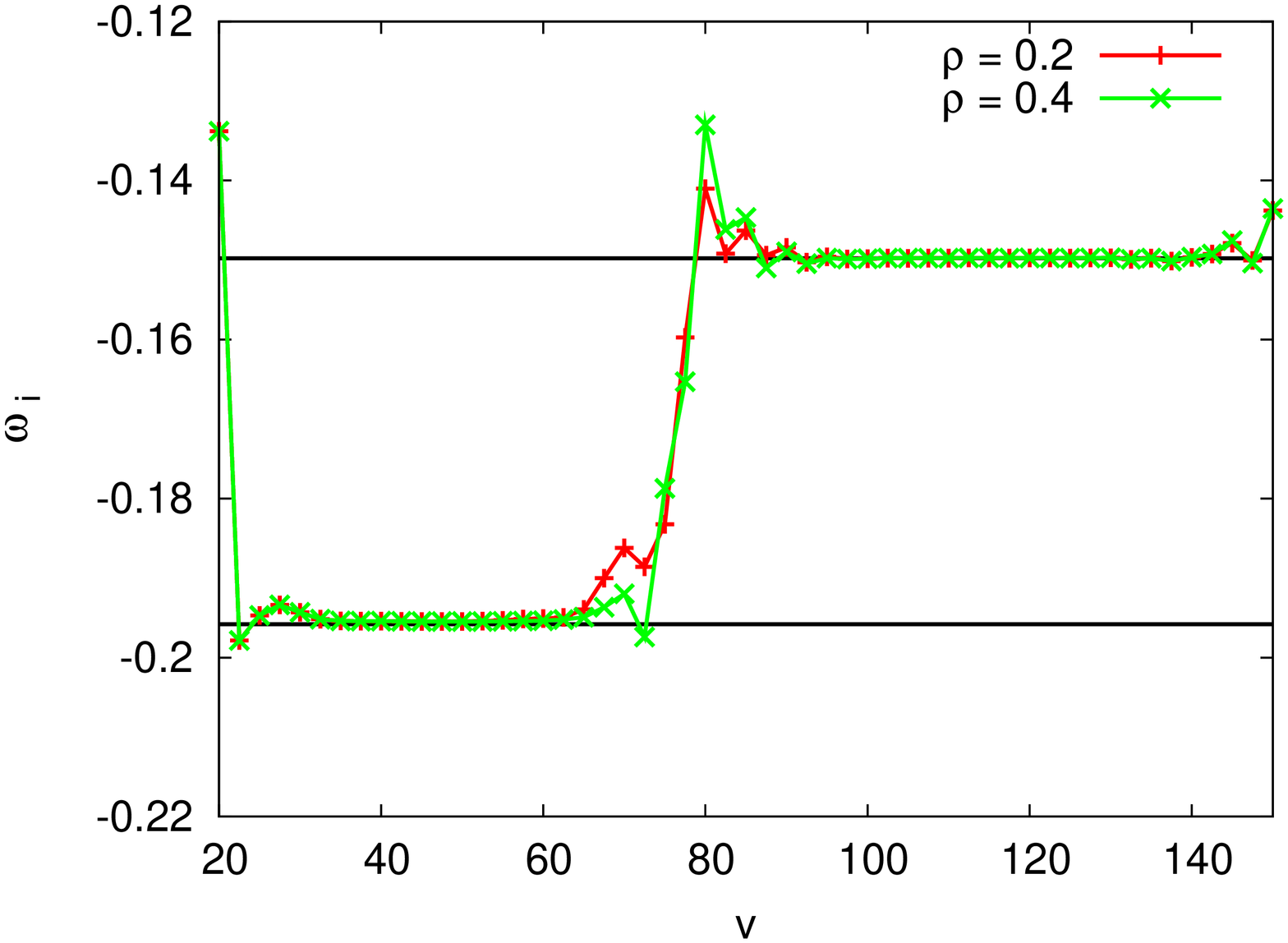}
  \includegraphics[angle=270,width=0.45\linewidth]{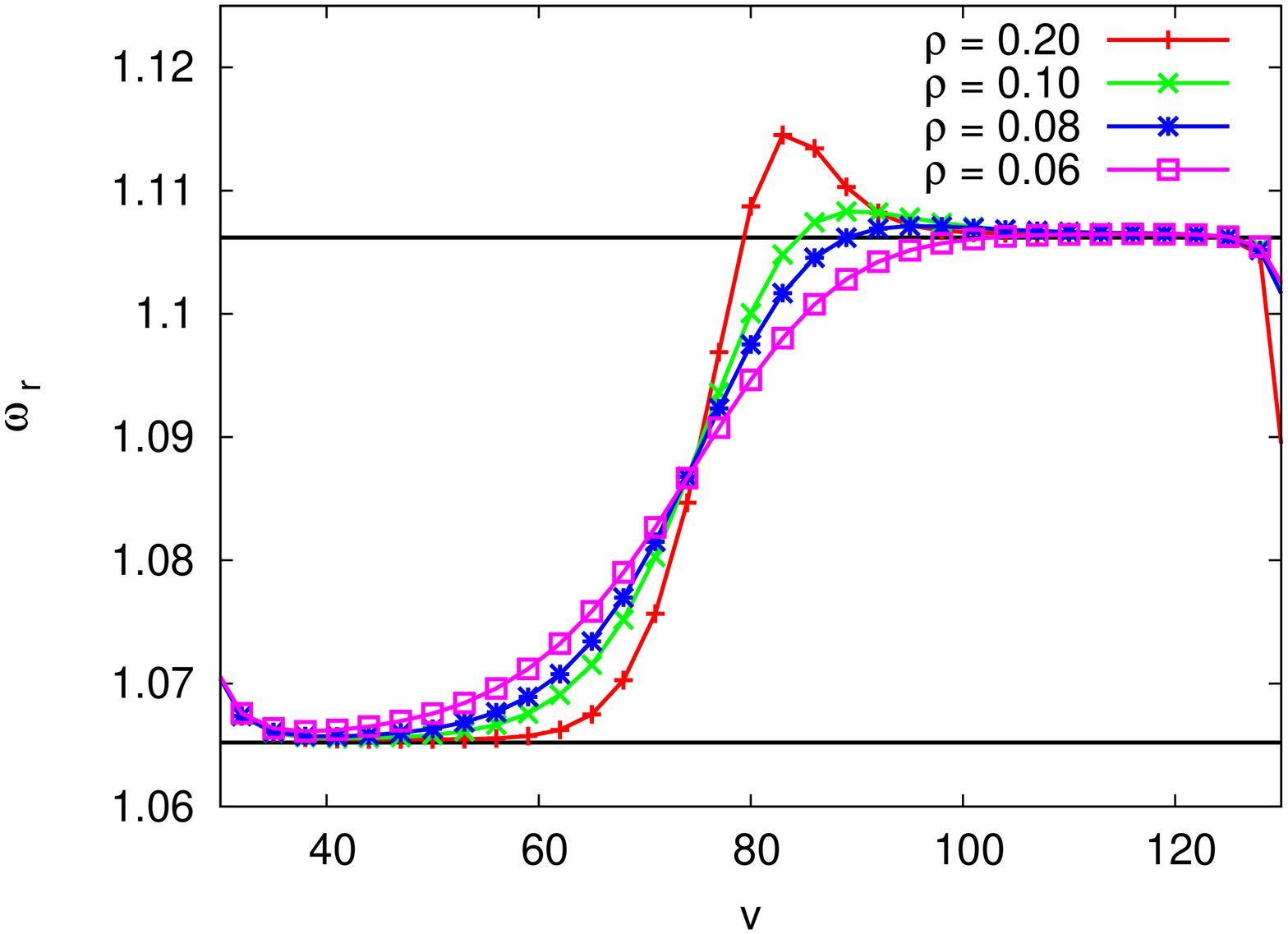}
  \includegraphics[angle=270,width=0.45\linewidth]{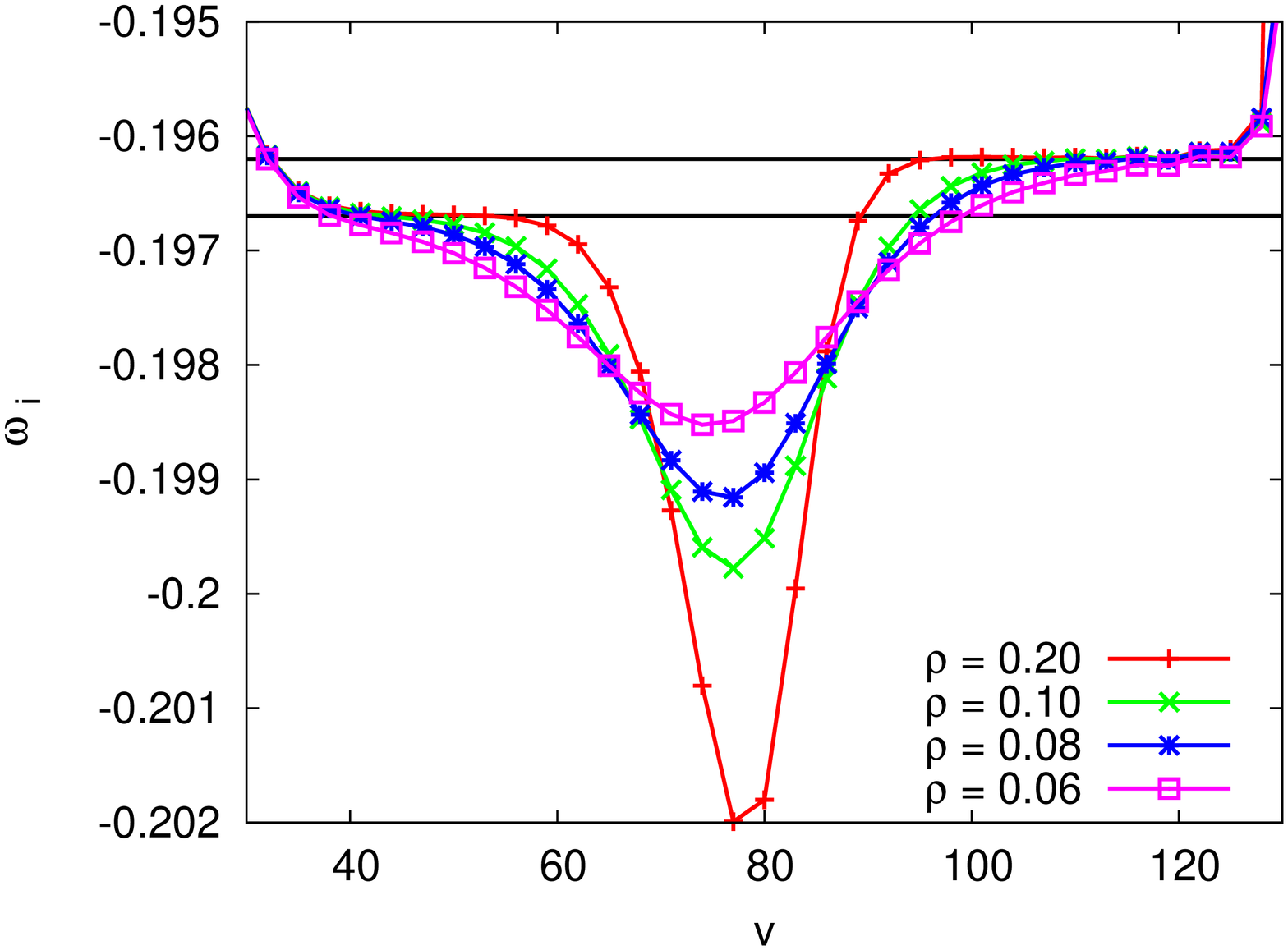}
\caption{Upper plot: $\omega_r$ (left) and $\omega_i$ (right) as
  function of $v$ for the $\ell = 2$ scalar perturbation of a black
  hole with constant charge $q = 0.25$ and $m(v)$ given by
  (\ref{tanh}) with $m_i = 0.5$, $m_f = 0.65$, $v_m = 75$ and
  different values of $\rho_m$. Lower plot: $\omega_r$ (left) and
  $\omega_i$ (right) as function of $v$ for the $\ell = 2$ scalar
  perturbation of a black hole with constant mass $m = 0.5$ and $q(v)$
  given by (\ref{tanh}) with $q_i = 0.35$, $q_f = 0.4$, $v_m = 75$ and
  different values of $\rho_q$. In all graphics, the horizontal lines
  show the frequency values for static RN black holes with the initial
  and final configurations. The fit for the frequencies is typically
  more sensitive for the values of $\omega_i$.}
\label{fig_mv}
\end{figure*}

In Fig.~\ref{fig_mv},
we present the time variation for the real and imaginary parts of the
QNM of the scalar perturbations with $\ell = 2$ for a 4-dimensional RN
black hole with   time dependent mass and charge functions. The QNM
frequencies change in a similar way with the time variation of the
mass or the charge, but they are typically more sensitive to the
charge variations.

\begin{figure*}[!htb]
  \includegraphics[angle=270,width=0.45\linewidth]{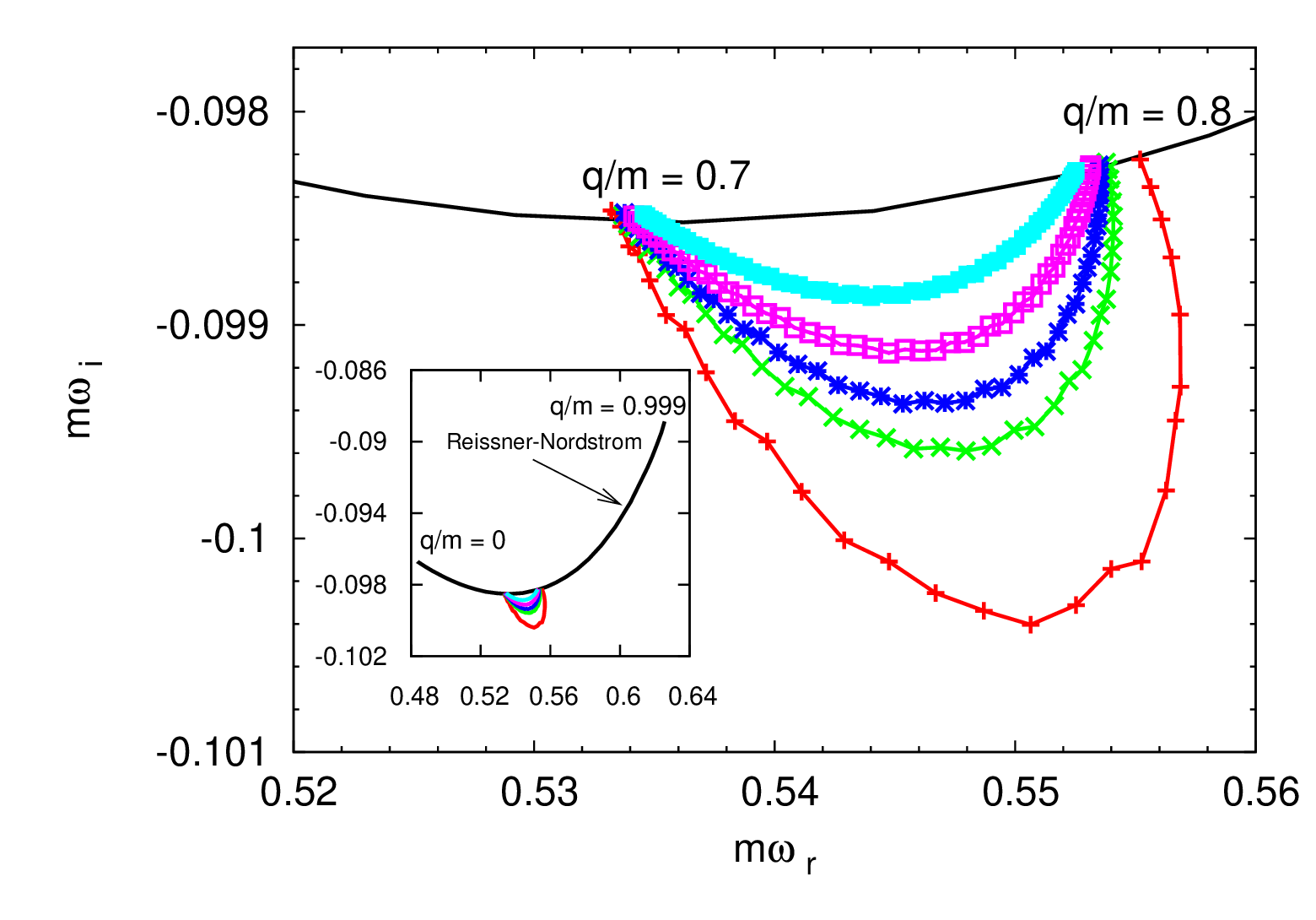}
  \includegraphics[angle=270,width=0.45\linewidth]{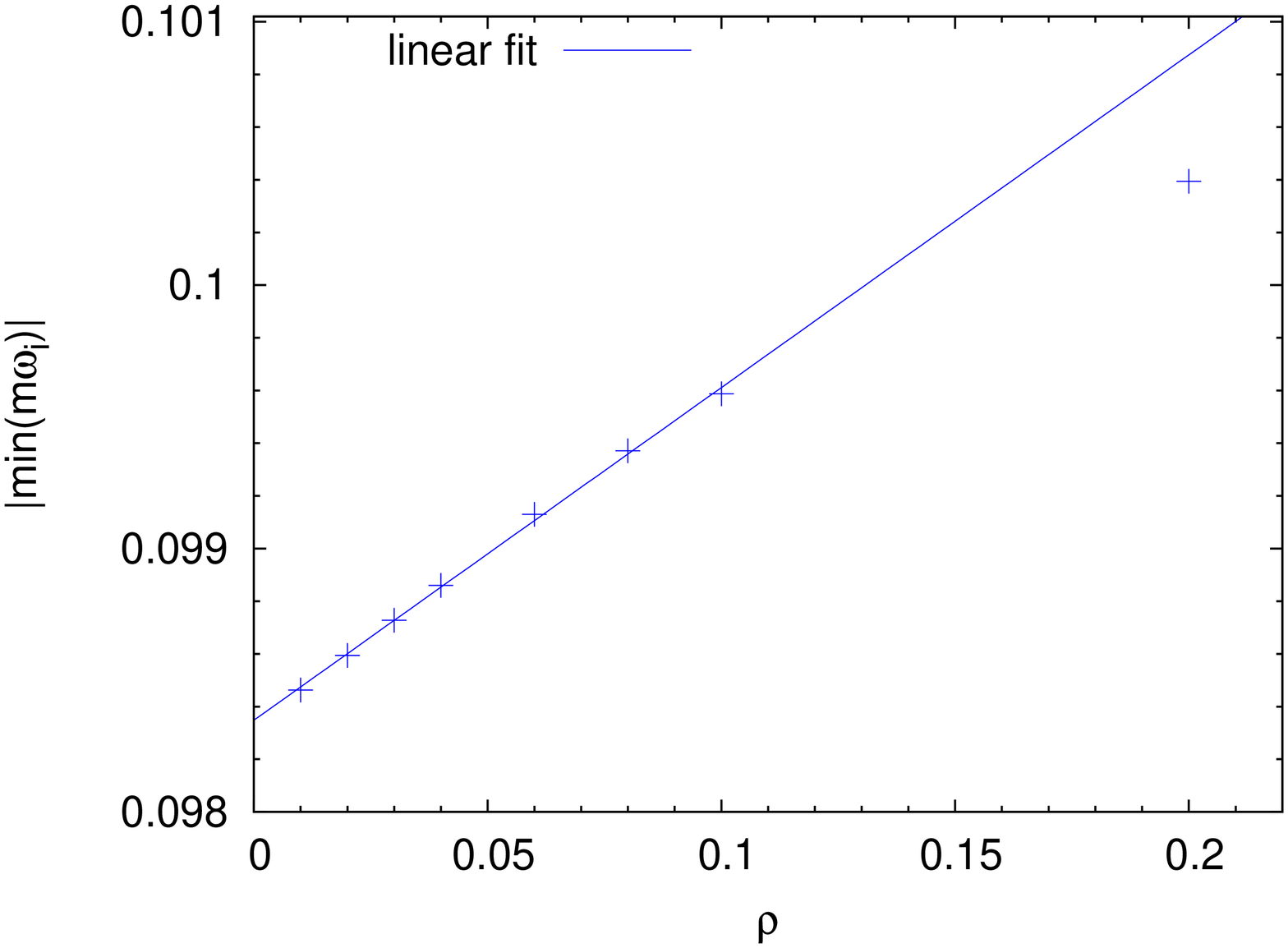}
\caption{Left plot: $\omega_r \times \omega_i$ plane, showing the
  transition between equilibrium states for a varying charge case
  according to (\ref{tanh}). The values of $\rho_q$ for these curves
  are 0.04, 0.06, 0.08, 0.1 and 0.2 (from top to bottom). Right plot:
  the minima of $\omega_i$ obtained in the transition shown in the
  left plot for different values of $\rho_q$, to quantify the
  non-adiabatic behavior.}
\label{wrxwi_01}
\end{figure*}

The results from the lower plot of Fig.~\ref{fig_mv}
can also be seen in Fig.~(\ref{wrxwi_01}),
in the $\omega_r \times \omega_i$ plane, where we see how the
variation goes from one stationary  state to another through a
nonstationary trajectory. In the right plot of Fig.\ref{wrxwi_01}, we
use the minimum value of the variation of $\omega_i$ in the
nonstationary trajectories of the left plot to quantify the deviation
from the stationary  trajectory. These values are plotted against
$\rho$, which can be related to a measure of how fast the variations
are. We can see that the behavior can be described extremely well by a
linear fit at first, deviating from this fit for $\rho \gtrsim 0.1$
(faster variations).

\begin{figure*}[!htb]
  \includegraphics[angle=270,width=0.45\linewidth]{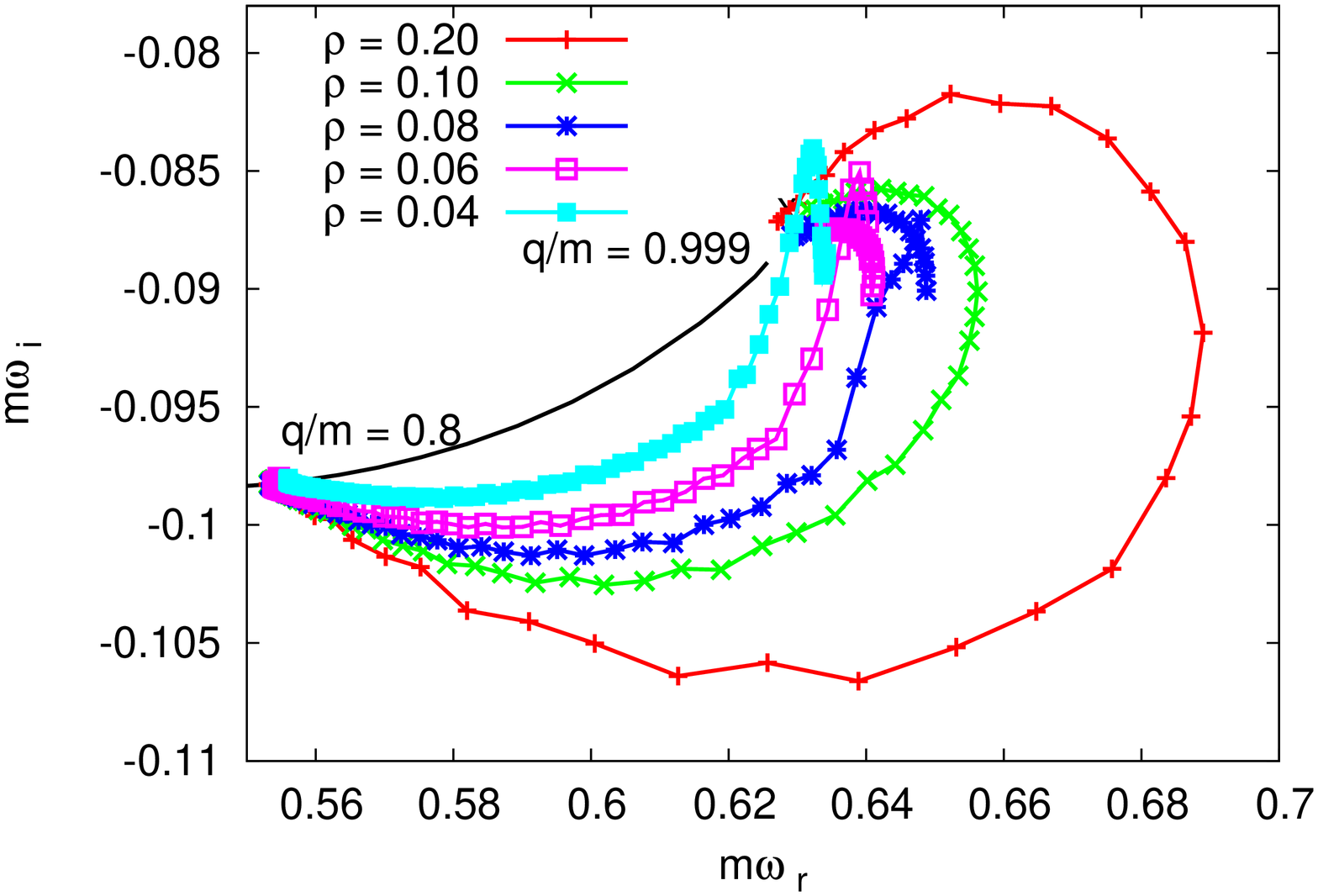}
  \includegraphics[angle=270,width=0.45\linewidth]{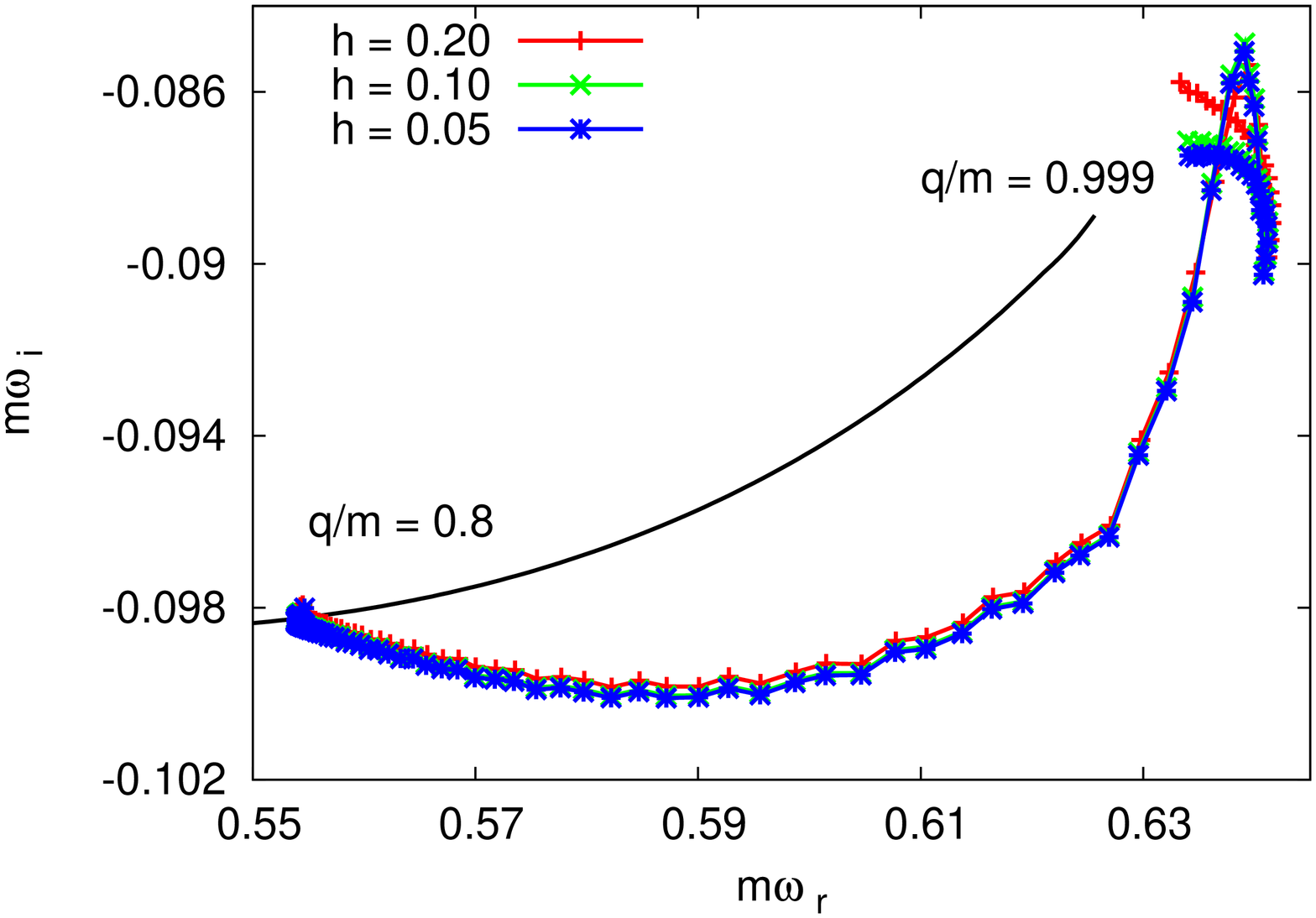}
\caption{Left plot: $\omega_r \times \omega_i$ plane, showing the
  transition between equilibrium states for a varying charge case
  according to (\ref{tanh}). The final state is an extremal RN black
  hole $(q=m)$ and the values of $\rho_q$ are the same of
  Fig. \ref{wrxwi_01}. Right plot: Convergence test made for a fixed
  value of $\rho_q=0.04$ and three different resolutions obtained with
  varying values for the integration stepsize $h$.}
\label{wrxwi_03}
\end{figure*}

The case of asymptotically extremal $(q_f=m_f)$ black holes is
specially notable in our approach due to numerical
technicalities. Our results for this case are presented in Fig.~\ref{wrxwi_03}.
In the right plot  we present a convergence test, essential in this
case, with the results obtained for different resolutions, while in
the left plot we show how the results depend on the manner in which
$q(v) \to m$. We remind here that an extremal RN black hole is, of
course, yet a regular black hole and not a naked singularity, and its
QNMs are well defined. It is only due to numerical limitations that
most works in the literature do not reach this limit.

The left plot of Fig.~\ref{wrxwi_03} shows once again that the deviation from the stationary behavior increases as we consider faster variations of the background, which are quantified by $\rho$. As this deviation becomes smaller, however, it is possible to note a nontrivial trajectory of the complex frequencies in the plane (more easily discernible in the right plot), whose origin is still unclear.

\subsection{Formation of a naked singularity}

We can use our formalism and numerical setup to probe the formation of
a naked singularity in the spacetime. The idea that it would be
possible to have regular scattering from naked singularities is rather
old \cite{Sandberg:1975sj}. In fact, as it was first noticed by
Gibbons \cite{Gibbons:1975jb}, minimally coupled scalar fields have
the remarkable property of being regular at the origin of a
Reissner-Nordstr\"om solution, where the spacetime manifold is
irremediably singular. The non-singular behavior of scalar
\cite{Martellini:1978jz} and other \cite{Belgiorno:1997wk} fields
around a Reissner-Nordstr\"om naked singularity has been investigated
since then. Despite being mathematically well posed, the field
dynamics near a naked singularity are typically ambiguous from the
physical point of view since it is not clear which boundary condition
one needs to impose for the field at the singularity.

The main motivation of this part of our investigation is the work of
Ishibashi and Hosoya \cite{Ishibashi:1999vw}, which  demonstrated
explicitly that is indeed possible to have unambiguous and  regular
scattering from naked singularities. After all, the presence of a
naked singularity might be less harmful than originally
conceived. This has been explored recently in a quantum field theory
scenario in the series of papers
\cite{Pitelli:2008pa,Pitelli:2009kd,Letelier:2010pv,Letelier:2010zz}.

We apply our numerical setup to simulate a charged black hole losing
mass  while keeping its charge constant, leading to a
Reissner-Nordstr\"om naked singularity.
\begin{figure*}[!htb]
  \includegraphics[angle=270,width=0.45\linewidth]{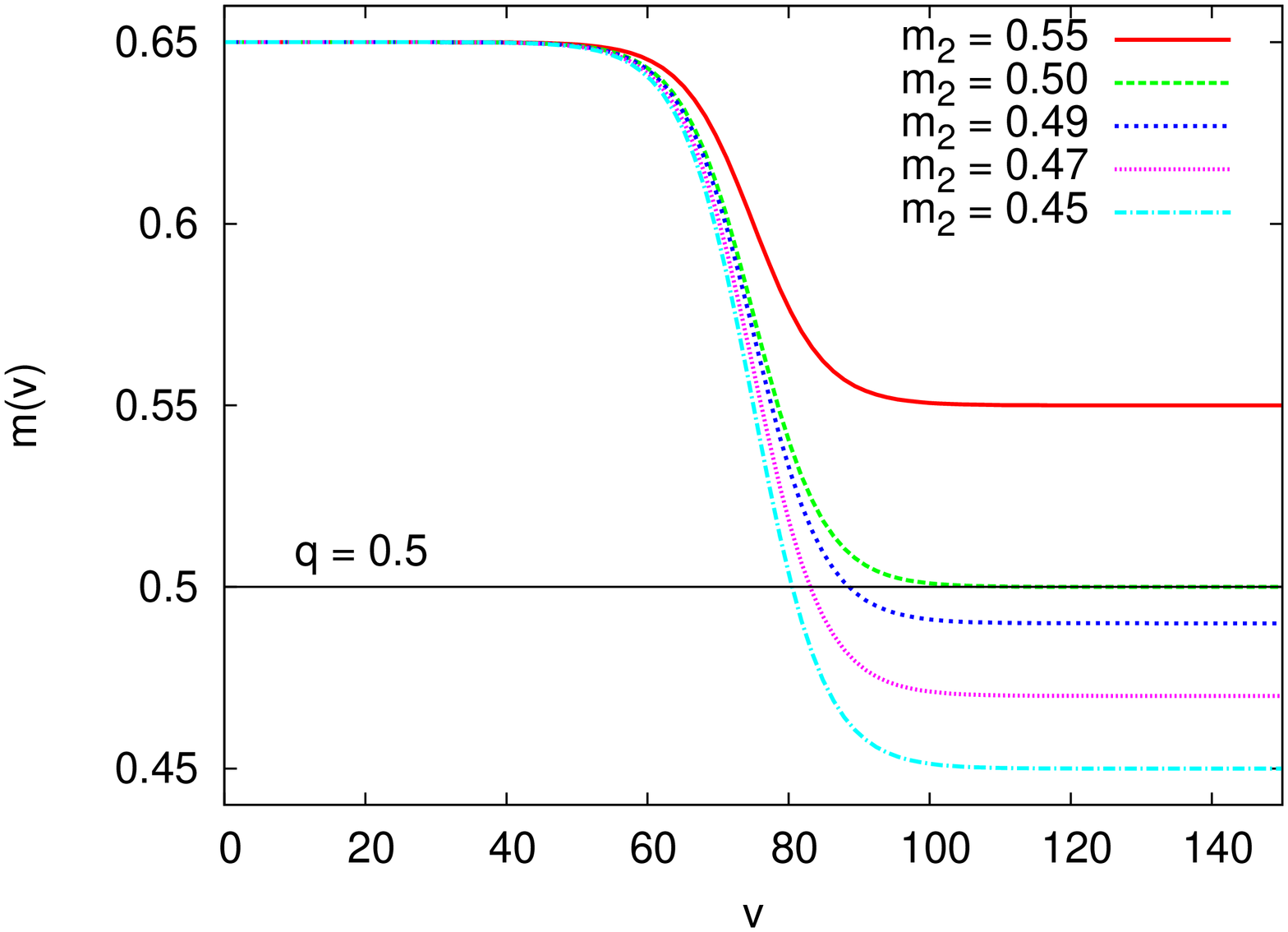}
  \includegraphics[angle=270,width=0.45\linewidth]{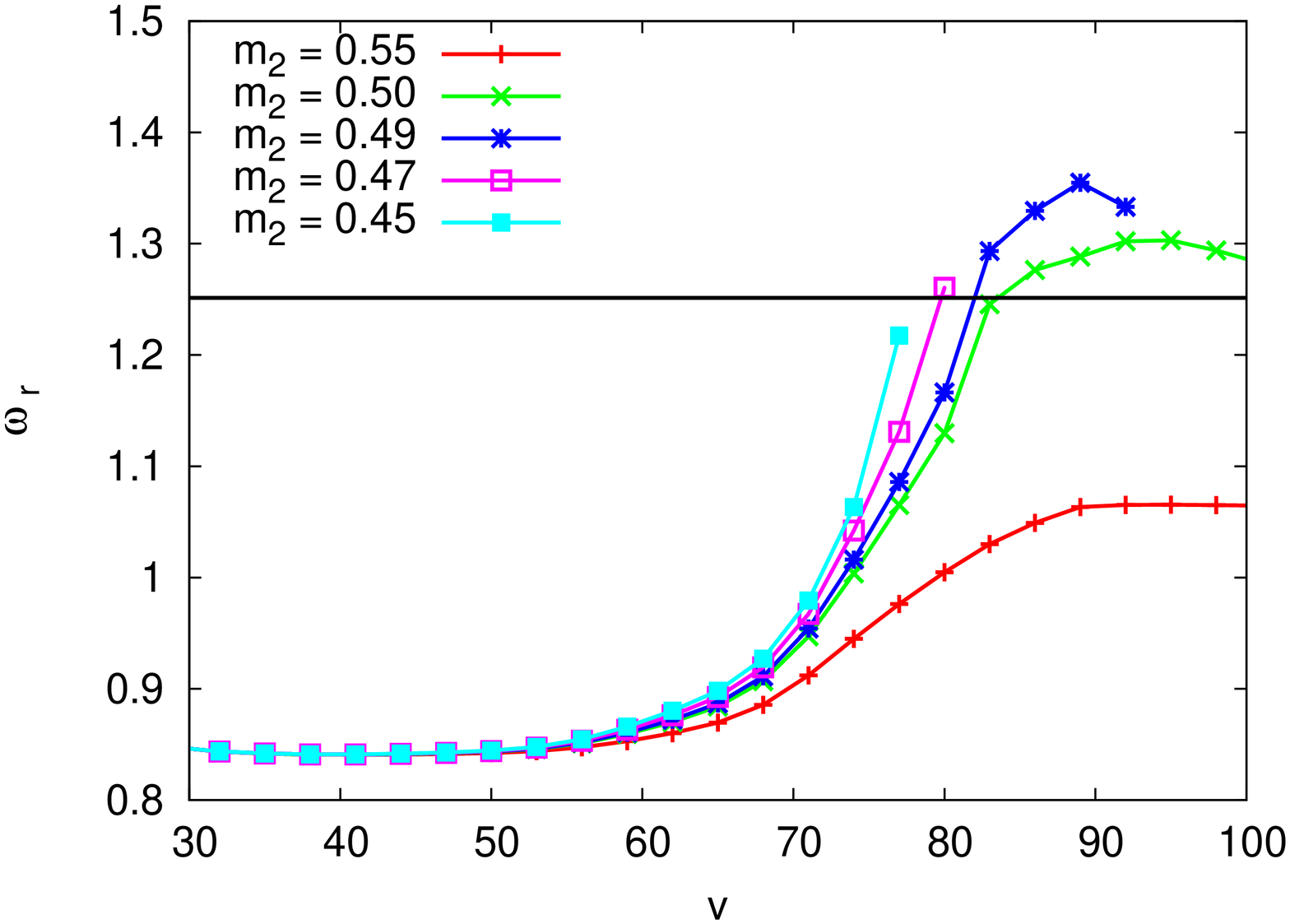}
\caption{Left plot: mass functions of the form
(\ref{tanh}) with $m_i = 0.65$, $\rho_m = 0.1$, $v_m = 75$ and
  different values of $m_j$, used to force the formation of a naked
  singularity while keeping $q = 0.5$ constant. Right plot: $\omega_r$
  as function of $v$ for the $\ell =  2$ scalar perturbation of black
  holes with constant charge and varying masses as given in the left
  plot. The horizontal line shows the approximate frequency for an
  extremal RN black hole. The behavior of $\omega_i$ is quite similar
  but typically known with less precision.}
\label{mc_of_v}
\end{figure*}
For a fixed charge value $q$, we set a mass function (\ref{tanh})
between two values $m_i$ and $m_f$, and take the final mass to be $m_f
< q$. At the point $v_*$ when $m(v_*) = q$ the evaluation of
$\varphi(u,v_*)$ is problematic since for the case of a naked
singularity the full range of $u$ must include the singularity at
$r=0$. Despite the fact that the scalar field can indeed be finite at
$r=0$ \cite{Martellini:1978jz}, our code cannot deal properly with the
terms like $V(u,v)\varphi(u,v)$ in the vicinity of the
singularity. This point is now under investigation, and a better
numerical code is being developed for this case.

Our objective here is
to investigate how the QNM behave until the very last moment before the
formation of a naked singularity. The results obtained for the
$\omega_r$ part of the QNMs are presented in Fig.~(\ref{mc_of_v}).
All simulations end at the point when an extremal RN black hole is formed, but the frequencies are all different! This is the result of the different nonstationary trajectories followed by the QNM in each case. We remark here that in the cases with a more rapid variation $m'(v_*)$ it seems there is not enough time before the simulation stops to see the maximum deviation of the frequency and the subsequent relaxation to a possible asymptotic value.

\section{Final remarks}

Our results confirm, for the general Vaidya case, the same
nonstationary behavior corresponding to the inertia of the QNM
frequencies  identified in \cite{Abdalla:2006vb}. In particular, for
situations with $r''_+(v) > \omega_i (v)$, the QNM frequencies will
not follow in the $\omega_r \times \omega_i$ plane a trajectory
corresponding to the instantaneous frequency associated to a RN black
hole of a given $q/m$ ratio (see Figs.~\ref{wrxwi_01} and
\ref{wrxwi_03}), and the inertial behavior of the QNM can be
identified. Moreover, we see that the faster the change in $m(v)$ or
$q(v)$ is, the bigger the deviation from the stationary regime will
be. We could determine that the deviation from the stationary behavior
is proportional to $\rho$ for masses and charges varying according to
(\ref{tanh}), at least for values of $\rho\lesssim 0.1$, see
Fig. \ref{wrxwi_01}.

Similarly to the uncharged case \cite{Abdalla:2006vb}, the imaginary parts of
the QNM frequencies are typically more sensitive to mass and charge
changes than the real parts.
However, we have  identified a new
behavior whose physical origin is still unclear for us: the QNM
frequencies are typically more sensitive to charge than mass
variations (compare the upper and lower graphics in
Fig. \ref{fig_mv}).

This leads to a curious fact and a further conclusion. Consider two
situations such that, in the first, electric charge is constant and mass is
increasing, let us say according to (\ref{tanh}). In the second, mass
is constant and electric charge is varying in a way that both cases
have the same quotient $q/m$ as function of $v$. For these two
situations, the instantaneous  QNM frequencies will  evolve
differently in the $\omega_r\times \omega_i$. According to our results, the case of
varying charge will typically show a bigger deviation from the
stationary curve $q/m$.
This new behavior demonstrates explicitly that
the ratio $q/m$ is not enough to characterize a non-stationary black
hole. (This does not represent any challenge to the no-hair theorems, though,
since they typically deal with stationary solutions.)

As for the $q=0$ case considered in \cite{Abdalla:2006vb}, an
interesting extension of this work would be analysis of the highly
damped QNM. Since for such overtones the
ratio $|\omega_i/\omega_r|$ is always larger than for the fundamental
($n = 0$) QNM considered here, including, for sufficient large $n$,
cases for which $|\omega_i/\omega_r|>1$, it would be interesting to
check if the stationary behavior could be improved for $n > 0$. In
particular, it would be very interesting to test the high sensitivity
of the instantaneous QNM overtones to electric charge variations. We
remark that the numerical analysis presented here cannot be
directly extended to the $n > 0$ case since one cannot identify the overtones
numerically with sufficient accuracy. We believe, however, this could
be attained by means of the WKB approximation.

\acknowledgments

This work was supported by CNPq, FAPESP and the Max Planck Society. It
is a pleasure to thank Rodrigo Panosso Macedo and Luciano Rezzolla for
useful discussions.

\end{document}